# *InterPhon*: *Ab initio* Interface Phonon Calculations within a 3D Electronic Structure Framework

*In Won Yeu[a], Gyuseung Han[a,b], Kun Hee Ye[a,b], Cheol Seong Hwang[b], and Jung-Hae Choi[a,\*]*

[a]Electronic Materials Research Center, Korea Institute of Science and Technology, Seoul 02792, Korea

[b]Department of Materials Science and Engineering, and Inter-University Semiconductor Research Center, Seoul National University, Seoul 08826, Korea

[\*]Corresponding author. Tel.: +82 2 958 5488; Fax: +82 2 958 6658

[\*]E-mail address: choijh@kist.re.kr (J.-H. Choi)

# Abstract

This work provides the community with an easily executable open-source Python package designed to automize the evaluation of Interfacial Phonons (***InterPhon***). Its strategy of arbitrarily defining the interfacial region and periodicity alleviates the excessive computational cost in applying *ab initio* phonon calculations to interfaces and enables efficient extraction of interfacial phonons. ***InterPhon*** makes it possible to apply all of the phonon-based predictions that have been available for bulk systems, to interfacial systems. The first example, in which this package was applied to InAs surfaces, demonstrates a systematic structure search for unexplored surface reconstructions, navigated by the imaginary mode of surface phonons. It eventually explains the anisotropic surface vibrations of the polar crystal. The second example, involving oxygen adsorption on Cu, reveals adsorption-induced vibrational change and its contribution to energetic stability. The third example, on a Si/GaAs interface, shows distinct vibrational patterns depending on interfacial structures. It leads to a prediction regarding the structural transition of interfaces and unveils the processing conditions for spontaneous growth of GaAs nanowires on Si. High-level automation in ***InterPhon*** will be of great help in elucidating interfacial atomic dynamics and in implementing an automated computational workflow for diverse interfacial systems.

# Program Summary

*Program title:* InterPhon

*Developer's repository link:* https://github.com/inwonyeu/interphon

*Licensing provisions:* LGPLv2.1

*Programming language:* Python

*Supplementary material:* a PDF file descrbing details of phonon formalism using FDM, validation of symmetry functionality, *InterPhon* package architecture, method to define the interfacial region and convergence test, and calculation details for all of the results in this work.

*Nature of problem:* The interface possesses diverse atomic structures and lattice vibrations, which are distinct from the bulk. In particular, interfacial phonons play the key roles to unveil the largely unexplored atomic dynamics within the localized region, and this information is essential to make a prediction regarding the dependence of interface structures on process conditions. However, there has been a limitation in applying *ab initio* phonon calculations to interfaces due to the excessive computational cost, introduced by their large number of atoms and broken symmetry. The problems are intrinsically inevitable within a three-dimensional (3D) DFT framework representing interfacial systems by supercells.

*Solution method:* Although the main obstacles are unavoidable, distinct interfacial phonons are confined to the vicinity of the interface. By limiting the range of phonon calculations to user-defined interfacial region, the enormous computational cost is mitigated. The strategy is efficiently implemented in a Python library capable of calculation setup, evaluation, analysis, and visualization for arbitrary interfacial systems in conjunction with any 3D DFT code. All of the functionality is fully automated and the program execution can be managed through high-level user interfaces without difficulty.

***Additional comments:*** At the time of writing, the latest version of *InterPhon* is 1.1.0, in which the conjunction with VASP, Quantum ESPRESSO, and FHI-aims is supported.

# 1. Introduction

The computation of collective atomic vibrations, called phonons, is a key to predicting a variety of material properties: transport properties such as diffusivity and conductivity; thermal properties such as heat capacity and entropy; and structural properties such as equilibrium phase and transition pathways [1]. With the aid of tremendous advancements in both hardware (high-performance parallel computers) and software (density functional theory (DFT) codes), a large amount of the energy and force state of arbitrarily given materials, $E(\{R_I\})$ and $F(\{R_I\})$, where $\{R_I\}$ indicates the positions of constituent atoms, can now be concurrently computed using available codes of electronic structure calculations. In addition to the quantum mechanical derivation of energy and force, the advent of several phonon codes, including Phonopy [1], PHON [2], and others [3], has made *ab initio* phonon calculations routine subjects of research. Nowadays, *ab initio* phonon calculations are even incorporated as a step in sequences of automatic multistep workflows, eliminating tedious user intervention and manual management during each calculation step [4]. By combining such an automatic platform with a database or crystal structure prediction (CSP) algorithm, massive phonon calculations can be carried out within high-throughput screening approaches to discover unknown materials with optimal properties. In the entire workflow, the phonon calculations usually play a vital role in assessing structural stability [5,6].

However, all of the phonon computations mentioned above are restricted to bulk systems. This is because a great deal of computational cost (approximately one or two orders higher than that of conventional electronic relaxation) is required to execute the *ab initio* phonon calculations. The required number of computations is proportional to the number of atoms per unit cell, so lower cost is expected for smaller bulk systems. A large number of symmetry operations of bulk crystals also provide an opportunity to substantially reduce the required number of computations [1,2]. In contrast, the symmetrically reduced and thus viable computational cost does not apply to defective systems with broken symmetry, such as surface and interfacial systems. A defective region introduces different kinds of vibrations originating from the different stoichiometry and bonding geometry compared to the bulk environment [7,8]. When performing defect calculations within three-dimensional (3D) DFT codes, however, the supercells of slab geometry or those with confined point defects must be generated [9,10] in order to remove any fictitious interactions between periodic cells [11,12]. Accordingly, a large number of atoms and a low degree of symmetry are intrinsically inevitable, which impose enormous computational cost. This cost is normally not affordable; thus, the effects of localized vibrations are commonly neglected in *ab initio* calculations for defective systems [13–16]. The accuracy of the results relies on the assumption that the vibrational contributions to the total free energy are not critical and only the electronic contributions are sufficient, which may not necessarily be the case.

Recently, more and more reports have emphasized the importance of the vibrations in interfacial (including surface) regions, for example, the atomistic mechanisms of a temperature-driven change in metal-insulator surface structure [17–19], equilibrium crystal shape [20], and nanowire growth behavior [21,22]. All of these, which were not explicable using only electronic energy contributions, were theoretically elucidated by looking at the subtle interplay between electronic and vibrational energies. Moreover, the emergent phenomena at interfaces [23–26], such as the interfacial phase-change memory (iPCM) [27] and multiferroics [28,29], are highly related to the intriguing inter-coupling between electronic and atomistic (vibrational) structures within the localized interfacial region. Therefore, the ability to compute interfacial phonon behavior will pave the way for a great deal of further research, and the development of a cost-effective and easily executable phonon code is timely.

In this work, an open-source Python package for interfacial phonon calculations, called *InterPhon*, is reported. It characterizes the distinct lattice vibrations with reduced periodicity (one-dimensional (1D) or two-dimensional (2D)) by automatically processing the information obtained by 3D-based DFT codes. It fully implements a set of tools to promote efficient phonon calculations and analyses for arbitrary interfacial systems in reduced symmetry and periodicity. Individuals with access to existing DFT programs can easily utilize the *InterPhon* package with the help of extensible compatibility supported by its adapter modules. The only requirements for package execution are a DFT input file representing the structure of interest and the corresponding output files providing the forces acting on each atom.

To demonstrate the usefulness of *InterPhon* and its applicability to a variety of material systems, three examples of package usage are presented. The examples investigate changes in interfacial phonon behavior that are associated with interfacial bond formation. They are surface reconstruction of InAs(111)A and (111)B, adsorption of oxygen on Cu(111), and interface of (111)-oriented Si/GaAs. The investigations demonstrate that *InterPhon* makes it possible to apply all phonon-based investigations that have been available for bulk systems, such as structure search, assessment of structural stability, and temperature-driven phase transition, to interfacial systems. Such analyses have hardly been achieved due to the scarcity of a cost-effective and easily executable software for interfacial phonons. Since the reconstruction, adsorption, and interface calculations are the general basis of *ab initio* modeling of material interactions, the *InterPhon* code will be valuable for designing diverse interactions by enabling predictions of interfacial dynamics at the atomic scale.

## 2. *InterPhon* background

### 2.1. Phonon computation

A phonon calculation starts with transforming a displacement vector field defined over lattice domain into a numerical problem of eigenvalues (phonon frequencies) and eigenvectors (phonon modes) of the dynamical matrix [30]. There are essentially two ways to evaluate the dynamical matrix: the direct approach using the finite displacement method (FDM) [31,32] and the linear-response approach using density functional perturbation theory (DFPT) [33,34]. *InterPhon* is based on the direct approach within the FDM scheme; the specific formalism used to evaluate the interatomic force constants (Equation S4) and dynamical matrix (Equation S6) is summarized in Section S1 of Supporting Information. Note that the term 'unit cell' is used to refer to the cell of interest for phonon evaluation (*e.g.*, phonons of interface unit cell) throughout this work. When obtaining the force constant elements, atomic forces in response to a set of atomic displacements are required and the forces should be evaluated using supercell consisting of several unit cells. This is because the DFT forces calculated via the Hellmann-Feynman theorem are actually sums over forces between one atom and all periodic images of another [31]. Assuming that atomic interactions for separation larger than supercell size (which eventually determines cutoff distance) are zero, the periodic contribution is neglected and the given DFT forces can be treated as the forces induced by the two-atom interaction. Then, considering the original periodicity, the force constant matrix of supercells is merged back into the unit cell dynamical matrix (*i.e.*, lattice-sum over $l'$ in Equation S6), which encompasses long-range atomic interactions.

## 2.2. *InterPhon* strategy: selection of the interfacial atoms

The direct approach within the FDM is also the operating mechanism behind widely used phonon codes for 3D bulk systems, such as Phonopy [1] and PHON [2]. These conventional codes, which take advantage of the symmetry in 3D space groups, generate the dynamical matrix of all constituent atoms in a unit cell by requiring 'complete FDM' for all atoms. By contrast, to make the interfacial phonon calculation feasible, ***InterPhon***, which requires 'selective FDM', generates the dynamical matrix of a subset of the constituent atoms in a unit cell and does not assume 3D periodicity. In conventional *ab initio* interface calculations, the excessive number of atoms is inevitable to represent interfacial region in a 3D periodic cell. Among all of these atoms, however, the bonding environment of most of the atoms located far from the interface (gray atoms in Fig. 1a) is similar to that of the bulk atoms, presenting the same vibrations as the ideal bulk (as intensively tested in Section S4.2 of Supporting Information). Deviation of vibrations from bulk occurs only on the atoms in the vicinity of the interface (green and orange atoms in Fig. 1a).

Therefore, ***InterPhon*** focuses on the interfacial atoms by allowing users to easily select atoms to be considered as the interface. In addition, contrary to the compulsory setting limited to the 3D periodic boundary condition (PBC) in existing DFT and phonon codes, user-defined setting of the PBC is allowed in ***InterPhon***. As a result, phonon evaluation proceeds only in the interfacial region with reduced PBC dimensions.

The key strategy behind ***InterPhon***, the user-defined setting of the interfacial region and PBC, reduces the cost of the most time-consuming element of phonon computations (that is, DFT force calculations for displaced supercells) by approximately one order of magnitude. This makes the calculation loads affordable. Eventually, the operation architecture, which is fundamentally different from that of the phonon codes for 3D bulk systems, allows efficient extraction of the localized phonons at arbitrary reciprocal points (called k-points) in 1D or 2D Brillouin zone (BZ). In addition, focusing on one of two exposed interfaces (top and bottom in slab geometry) makes it possible to evaluate phonons of a polar interface. Since the top and bottom interfaces along a crystallographically polar direction are inherently different, the phonons of an individual polar interface cannot be evaluated if phonon computation is performed for all the constituent atoms without limiting the interfacial region. Any nonlocal interaction between the bulk and interface layers is described during the stage of DFT electronic calculations prior to the execution of ***InterPhon***.

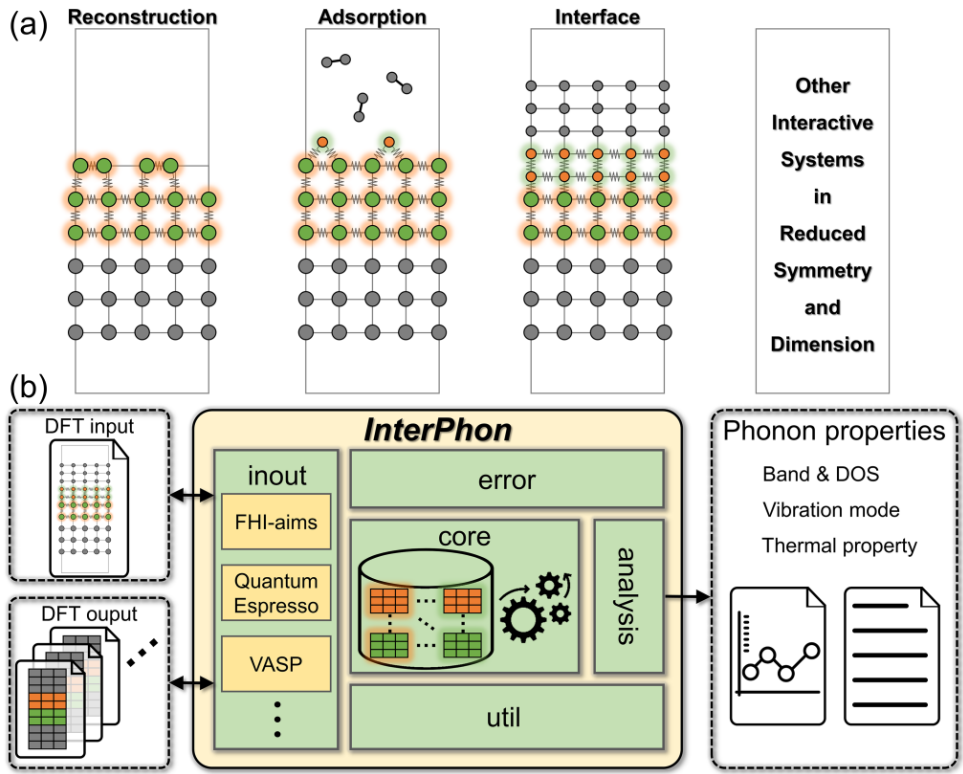

**Figure 1.** (a) Representative systems to which *InterPhon* can be applied. Phonon evaluation proceeds only on the selected atoms in the vicinity of the interface, which are shown in green and orange. The atoms embedded in bulk are shown in gray. (b) Schematic overview of *InterPhon* operation. The program execution starts with reading DFT input/output files in the 'inout' sub-package, followed by the numerical calculations in the 'core' sub-package, and ends with printing out the phonon properties by the 'analysis' sub-package. In the middle of the processes, the functionality supported in the 'util' and 'error' sub-packages are internally employed.

## 2.3. Symmetry functionality

The 2D symmetry of interfaces, although its number of operations is usually smaller than that in 3D crystal, enables further reduction of the number of DFT calculations required. In the following explanation on translational and point group symmetry, the notations summarized in Appendices B and C are used to avoid confusion, possibly caused by various representations of vector, matrix, and their product. By translational symmetry, translation by in-plane lattice vectors leaves the interface system unchanged. Therefore, the atomic displacements are only required to the atoms in one unit cell; total number of required displacements corresponds to the number of interfacial atoms in unit cell multiplied by 6, where 6 comes from the forward and backward displacements for each of the three Cartesian directions (see Equation S4). Note that the conventional translational invariance in the force constant matrix [2,31]:

$$\sum_{l's'\beta} K_{ls\alpha,l's'\beta} = 0, \tag{1}$$

which is derived from the fact that the atomic forces must be zero when a crystal shifts rigidly (*i.e.*, $F_{ls\alpha} = 0$ when $u_{l's'\beta}$ is constant for all the atoms in Equation B6), is usually not satisfied in interfacial phonon calculations with *InterPhon*. This is because unlike 2D materials such as graphene, the freely translational motions of interfaces are limited by the presence of bonds with neighboring bulk. Accordingly, acoustic phonon modes with zero frequencies at the Γ-point, corresponding to translational motions, will not be found in interfacial phonons as shown in the calculation results of this study.

In addition, the use of point group symmetry can reduce the number of atomic displacements even further. The displacement vectors do not necessarily have to be the exact Cartesian directions, but just needs to be three linearly independent vectors. If $u_{l's'1}$ is set and the other two linearly independent displacement vectors, $u_{l's'2}$ and $u_{l's'3}$ of the atom at $R_{l's'}$, can be obtained by applying point group symmetry operations $\{W\}$ of the interface plane group, displacement only along the one direction is sufficient for this atom [2,35]. After creating a set of three linearly independent displacement ($u_{l's'1}$, $u_{l's'2}$, $u_{l's'3}$) and resulting force vectors ($F_{ls,l's'1}$, $F_{ls,l's'2}$, $F_{ls,l's'3}$) for an two-atom pair ($R_{ls}$ and $R_{l's'}$), the force constant matrix in orthogonal Cartesian coordinates (Equation B4) can be recovered by:

$$K_{ls,l's'} = \begin{pmatrix} K_{lsx,l's'1} & K_{lsx,l's'2} & K_{lsx,l's'3} \\ K_{lsy,l's'1} & K_{lsy,l's'2} & K_{lsy,l's'3} \\ K_{lsz,l's'1} & K_{lsz,l's'2} & K_{lsz,l's'3} \end{pmatrix} \left( \frac{u_{l's'1}}{|u_{l's'1}|}, \frac{u_{l's'2}}{|u_{l's'2}|}, \frac{u_{l's'3}}{|u_{l's'3}|} \right)^{-1}. \qquad (2)$$

Furthermore, if a point group symmetry operation $W$ exists by which another two-atom pair ($R_{\widetilde{ls}}$ and $R_{\widetilde{l's'}}$) is mapped onto, the force constant matrix for the image atom pair does not need to be evaluated directly and is simply generated by:

$$W^{-1} K_{\widetilde{ls},\widetilde{l's'}} W = K_{ls,l's'}, \qquad (3)$$

which is derived in Appendix D.

In the backend of ***InterPhon***, this translational and point group symmetry mapping is conducted by automatically searching the in-plane symmetry operations of a given interface structure. Transformation candidates are chosen as symmetry operations if image points $\{R_{\widetilde{ls}}\}$ of all interfacial atoms are located at the atomic sites with the same atomic type. It should be noted that conventional 2D version of $R_{ls}$ (2×1 vector) and $W$ (2×2 matrix) causes serious algorithmic errors by allowing mapping between different normal coordinates. To avoid this possibility, in the example case that $a_1$ and $a_2$ lattice directions are aligned along the interface plane, 3×3 matrix $W$ and three coordinates of $R_{ls}$ are utilized when searching the in-plane symmetry operations by setting $W_{1,3} = W_{2,3} = W_{3,1} = W_{3,2} = 0$ and $W_{3,3} = 1$ in Equation C2. From the set of accepted symmetry operations, a 2D point group type (crystal class) is inferred by comparing with a set of determinant and trace of $W$, called look-up table [36]. This symmetry functionality is implemented as default in *InterPhon* and verified in Fig. S1 showing that applying symmetry operations enables a smaller amount of DFT calculations to provide the same phonon dispersion as the one from larger calculations without symmetry usage.

## 2.4. Architecture of *InterPhon*

*InterPhon*, which is steered by high-level user interfaces, automatically performs all of the processes required to characterize interfacial phonons from given DFT input and output files. The overall *InterPhon* package consists of five sub-packages: (i) 'inout', (ii) 'core', (iii) 'analysis', (iv) 'util', and (v) 'error'. Each of the sub-packages takes different roles and the main workflow is as follows: (i) modules in the 'inout' sub-package read and parser DFT input/output files; (ii) this information required to generate the dynamical matrix, such as crystal structure and atomic forces, is provided to modules in the 'core' sub-package that calculate phonon frequencies and modes; (iii) the processed information is provided to modules in the 'analysis' sub-package and the internal processes end with printing of the phonon properties, such as the density of states (DOS), band, vibrational motions (phonon mode), and thermal properties (vibrational entropy and free energy) in the complete forms of both data files and graphics (Fig. 1b). In the graphical representations, inspired by PyProcar [37], atom projection in the phonon band and DOS is effectively visualized in high quality to help analysis of phonon characteristics, using matplotlib [38] as a graphic library. In the middle of the processes, the modules in (iv) 'util' and (v) 'error' sub-packages are internally employed to enrich functionality. The detailed design architecture of the package and interrelation among the sub-modules are shown in Fig. S2, along with a full explanation of the operating procedures, in Section S3 of Supporting Information. The usage of parser modules in the 'inout' sub-package, which communicate with external files, allows a standardized representation of the internal data, irrespective of external DFT codes having different file formats. The parsers for VASP [39–41], Quantum ESPRESSO [42], and FHI-aims [43] are currently supported, and the parsers for other codes are under development.

### 2.5. Parameters of *InterPhon*

The key user-parameters of *InterPhon* are **displacement** (default = 0.02), **enlargement** (default = "2 2 1"), and **periodicity** (default = "1 1 0"). The default values mean that the displacement length for FDM (*i.e.*, *u* in Equation S4) is 0.02 Å; the extension ratios along the $a_1$, $a_2$, and $a_3$ lattice directions are 2, 2, and 1, which determine the size of supercells and accordingly define the cutoff range of the considered interatomic forces; and the periodic (1 or True) boundary conditions are along the $a_1$ and $a_2$ directions and open (0 or False) along the $a_3$ direction. According to the **periodicity** parameter, the interface plane direction, to which the in-plane symmetry functionality is carried out, is determined and even 0D (like isolated clusters) and 1D PBC systems can be handled. In addition, *InterPhon* defines the interfacial region according to the automatic reading of the statement of constraints on atom movements, which is supposed to be set by users within the DFT structure file (see Section S4.1 of Supporting Information). By providing users with the opportunity to define the interfacial region, the convergence of the phonon characteristics with respect to the increase in interfacial region thickness (*i.e.*, the number of selected atom layers) can easily be confirmed. Since the explicit boundary between the interfacial region and the bulk cannot be universally defined, this kind of convergence test is essential. An effective methodology to perform the convergence test is suggested in Section S4.2 of Supporting Information, and the convergence of all results in this work was tested. Except for the InAs surface reconstructions, where the bonding states change significantly, fast convergence in vibrations with respect to interfacial thickness was confirmed. This is a very encouraging result because it implies that further reduction of the interfacial region than was done in this work, where three layers were selected as the interfacial region, is possible, which would further decrease the computational cost.

## 2.6. Data and library availability

The following section presents the vibrational change introduced by interfacial bond formation, which was investigated by ***InterPhon*** in conjunction with VASP code. All of the data used in these investigations are available in the 'example' folder at the source code repository. The data provenance showing how the data were obtained can be traced back, enabling the reader to reproduce the results. The latest version of the ***InterPhon*** package has been released in the Github repository at https://github.com/inwonyeu/interphon, which can be easily installed via pip:

```
$ pip install interphon
```

In addition, a complete user manual on the installation, execution, file structure, and operation options is clearly described in the documentation at https://interphon.readthedocs.io. Following this documentation, ***InterPhon*** can be executed in two ways: command line and Python script (or interpreter). While the execution by command line is easy to follow, the Python script is more appropriate for dynamic utilization according to the user-specific purpose. Ensuring code reuse
15

and modularity of the Python language, the conjunction of compiled DFT codes with this object-oriented package will be an ideal approach to automated computation projects for heterogeneous interfacial systems [44,45].

## 3. Application examples of *InterPhon* and discussion

The results below show representative examples of *InterPhon* application: the surface reconstruction of InAs(111)A and (111)B, the adsorption of oxygen on Cu(111), and the interfacial structure of (111)-oriented Si/GaAs. Follwing these example investigations, *InterPhon* is applicable to other material systems for the predictions of temperature-dependent surface, adsorption, and interfacial structures, which will be crucial for elucidating the process-dependence of crystal growth, catalytic activity, and other correlated phenomena. The calculation details of how the VASP and *InterPhon* parameters were set for each example are given in Section S5 of Supporting Information.

### 3.1. Surface reconstruction: InAs(111)A and (111)B

The crystal structure of InAs is zinc-blende (ZB, $F\bar{4}3m$), which is non-centrosymmetric. Therefore, it contains various symmetrically-inequivalent surfaces, resulting in explicit anisotropy in the structural and physical properties. For instance, among the total of eight (111) surfaces, four (111) surfaces are terminated by cations, while the other four (111) surfaces are terminated by anions. These two types of inequivalent surfaces, which reside in opposite directions, are referred to as (111)A and (111)B, respectively. The intriguing asymmetric growth behavior of nanowires following these two opposite directions of <111> has recently been observed in ZB-structured materials [46,47]. The distinctive growth behavior, resulting from anisotropic interactions between vapor and solid surface, is eventually determined by the distinct surface reconstructions. In this section, the differences in surface structures and vibrations between InAs(111)A and (111)B are investigated to demonstrate how *InterPhon* can be applied to study the dynamic and thermal stability of surface reconstructions.

Figure 2 shows the surface reconstruction and phonon characteristics of the (111)A surface. The side view of the asymmetric slab composed of ten layers along the <111> direction in Fig. 2a shows that the bottom As-atoms at the (111)B surface were passivated by pseudo-hydrogen with 0.75 valence electrons. Figure 2b shows the top view of the (111)A surface and the process of creating the In-vacancy(2×2) reconstruction, which is known to be stable under most thermodynamic conditions [20]. After setting one vacant site on the topmost In-layer per (2×2) cell, a relaxation calculation (*i.e.*, optimizing the positions of constituent atoms into the energy minimum) was performed for the top five layers while the bottom five layers and the hydrogen atoms were fixed.

For this relaxed unit cell of the In-vacancy(2×2) reconstruction (solid parallelogram in the right panel of Fig. 2b), further FDM was carried out for the top three layers (a total of 11 atoms consisting of 3, 4, and 4 atoms in the top first, second, and third layer, respectively) by setting the user-parameters of *InterPhon* as follows: **displacement** = 0.02, **enlargement** = "1 1 1", and **periodicity** = "1 1 0" ($a_3$ direction corresponds to [111]). The corresponding phonon dispersion projected to the first layer (*i.e.*, three In-atoms) for 33 branches, *i.e.*, 3 (x, y, z directions) × 11 (selected atoms), is shown in Fig. 2c. The dispersion was obtained by setting one of the parameters of the 'plot' method in 'Band' class as **option** = 'projection' (for all of the available plotting options, see the online user manual). An imaginary mode is found at points $\bar{M}$ and $\bar{K}$, presumably due to the short cutoff range of interatomic forces [31] under the assumption that the force between atoms separated by more than one unit cell (whose cell size is $a_1 = a_2 = 8.5$ Å) is zero. Therefore, the interatomic forces were recalculated to incorporate the force interactions outside of the unit cell using the supercell doubled along the periodic directions, as represented by the dotted parallelogram in the right panel of Fig. 2b: **displacement** = 0.02, **enlargement** = "2 2 1", and **periodicity** = "1 1 0". The result shown in Fig. 2d does not have any imaginary mode, ensuring the dynamical stability of the In-vacancy(2×2). Figure 2d was obtained by utilizing the 'plot_with_dos' method of 'Band' class by setting some parameters as follows: **band_option** = 'projection' and **dos_option** = 'stack'. From the atom-projected DOSs, the relative contribution of each surface layer (first, second, and third) to the total DOS is easily identified: the low-frequency modes of the surface phonon are mainly contributed by three In-atoms in the first topmost layer.

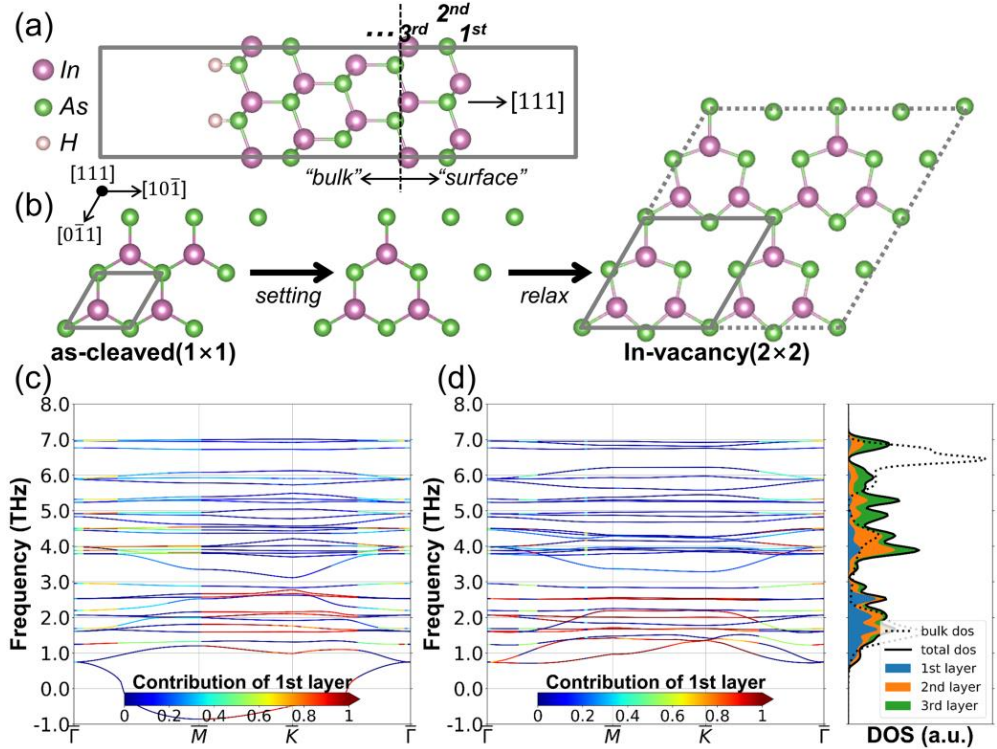

**Figure 2.** (a) Side view of the as-cleaved slab along the <111> direction of InAs. The bottom (111)B surface is passivated by hydrogens and the top three layers of the (111)A surface were selected for phonon evaluation. (b) Top view of the topmost bilayer (first In- and second As-layers) of (111)A as-cleaved(1×1) and In-vacancy(2×2). The solid parallelogram in the left panel indicates the unit cell of the as-cleaved(1×1). The solid and dotted parallelograms in the right panel indicate the unit cell of the In-vacancy(2×2) and the supercell accommodating four unit cells, respectively. The first-layer-projected phonon band of the In-vacancy(2×2) was evaluated using the DFT forces obtained within (c) the unit cell of the reconstruction (solid parallelogram), and (d) the supercell (dotted parallelogram), and are presented along with the layer-projected phonon DOSs in stack mode in comparison with bulk InAs (ZB, dotted line).

Figure 3 shows the surface reconstruction and phonon characteristics of the opposite surface, (111)B. The side view of asymmetric slab along the <111> direction in Fig. 3a shows that the bottom In-atoms of the (111)A surface were passivated by pseudo-hydrogen with 1.25 valence electrons. Figure 3b shows the top view of the (111)B surface and the process of creating the In-vacancy β(2×2) reconstruction, which was suggested to be stable under In-rich conditions [20]. After setting one vacant site on the topmost As-layer per (2×2) cell and substituting the remaining three As-atoms with In-atoms, a relaxation calculation was performed for the top five layers, resulting in the In-vacancy α(2×2) reconstruction. For the relaxed unit cell, further phonon analysis was carried out for the top three layers (a total of 11 atoms consisting of 3, 4, and 4 atoms in the first, second, and third layer, respectively) by setting the user-parameters of *InterPhon* as follows: **displacement** = 0.02, **enlargement** = "2 2 1", and **periodicity** = "1 1 0". The corresponding phonon dispersion projected to the first layer (*i.e.*, three In-atoms) for 33 branches is shown in Fig. 3c. Although the cutoff range of the interatomic forces was set to be the twice the length of the reconstruction unit cell (dotted parallelogram in Fig. 3b), which is large enough to incorporate long-range interactions, imaginary modes are found at all of the high-symmetry points, $\bar{\Gamma}$, $\bar{M}$, and $\bar{K}$. This implies that the In-vacancy α(2×2) reconstruction is dynamically unstable. The collective atomic motion corresponding to the lowest imaginary frequency at point $\bar{\Gamma}$ (indicated by a red circle and arrow in Fig. 3c) is indicated by the blue arrows on the topmost bilayer in Fig. 3b. Since the imaginary frequency (conventionally represented as negative real numbers) means that there is no restoring force against atomic motion along the corresponding imaginary mode, another relaxation calculation was conducted after slightly deforming the atomic positions along this imaginary mode.

As a result, the (111)B In-vacancy β(2×2) reconstruction was obtained, which is clearly distinguishable from the In-vacancy α(2×2) by the bonding geometry (bottom panel of Fig. 3b). One possible reason why the first relaxation converges to the In-vacancy α(2×2) is the constraint in the DFT energy minimization algorithm, which is caused by the energy landscape and initial symmetry. By escaping from the energy saddle point of the In-vacancy α(2×2) and changing to the In-vacancy β(2×2), the DFT electronic energy was lowered by 0.32 eV, corresponding to a decrease in surface energy by 5.1 meV/Å$^2$. This finding demonstrates that such an analysis of a surface-localized phonon provides a great opportunity to systematically search for stable surface structures, which is analogous to a systematic search of the stable bulk phase by 3D *ab initio* phonon calculations [48].

As the reconstruction changes, the surface phonon dispersion also changes. Figure 3d shows the first-layer-projected band along with the layer-projected DOSs of the (111)B In-vacancy β(2×2). Although all DFT and *InterPhon* calculation parameters were the same as those used for the In-vacancy α(2×2) in Fig. 3c, the imaginary modes disappear, showing the dynamical stability of the In-vacancy β(2×2). Furthermore, the phonon states contributed by the In-atoms in the first topmost layer are more localized to the high-frequency states in the In-vacancy β(2×2) compared to In-vacancy α(2×2), which is clearly shown by the color code presented by "contribution of 1st layer". On the other hand, the absence of a third layer contribution to the high-frequency vibrations (clearly shown in the layer-projected DOSs) implies that the atomic interactions between the In-atoms in the topmost bilayer and the As-atoms in the third layer become negligible in this reconstruction. This change in bonding states results in weaker bonding and substantially lower frequency than bulk InAs (ZB, dotted line in Fig. 3d).

Finally, the different stoichiometry and bonding geometry between the (111)A In-vacancy(2×2) and (111)B In-vacancy β(2×2) lead to substantial differences in surface phonons, as shown in Fig. 2d and Fig. 3d. Therefore, the corresponding vibrational free energy ($F^{vib}$), which can be evaluated by the integration of the phonon DOS (Equation S7) [1,2,49], would be highly anisotropic, as would the electronic energy. Figure 3e shows the electronic surface energy ($\gamma^{elec}$) and the total surface energy ($\gamma = \gamma^{elec} + \gamma^{vib}$) of the (111)A In-vacancy(2×2) and (111)B In-vacancy β(2×2) reconstructions as a function of temperature near the experimentally-relevant conditions for InAs growth by molecular beam epitaxy (As-vapor pressure of $10^{-9}$ atm). The vibrational surface energy ($\gamma^{vib}$) corresponds to the $F^{vib}$ difference between surface and bulk. The total surface energy ($\gamma$), containing the electronic and vibrational contributions, can be predicted as a function of temperature and vapor pressure by a method called *ab initio* thermodynamics that considers the contact-induced interactions between vapor environment and solid; the detailed method is fully described in method section of the authors' previous papers [21,22]. Comparing the (111)A and (111)B reconstructions, the different slope of $\gamma^{elec}$ is caused by differences in implicit temperature dependence, which comes from the atomic stoichiometry mismatch between surface and bulk [8]. Furthermore, it should be noted that the reduction in $\gamma$ by the contribution of $\gamma^{vib}$, called explicit temperature dependence, is much more significant in the (111)B reconstruction because of its substantially lower frequency compared to bulk InAs and the (111)A reconstruction (Fig. 2d and Fig. 3d). Since a majority of modern functional materials are made of polar crystals (like zinc-blende structures) [3], the ability to elucidate this kind of anisotropic surface vibration will be valuable for other scientifically and technologically important problems such as ferroelectrics and multiferroics.

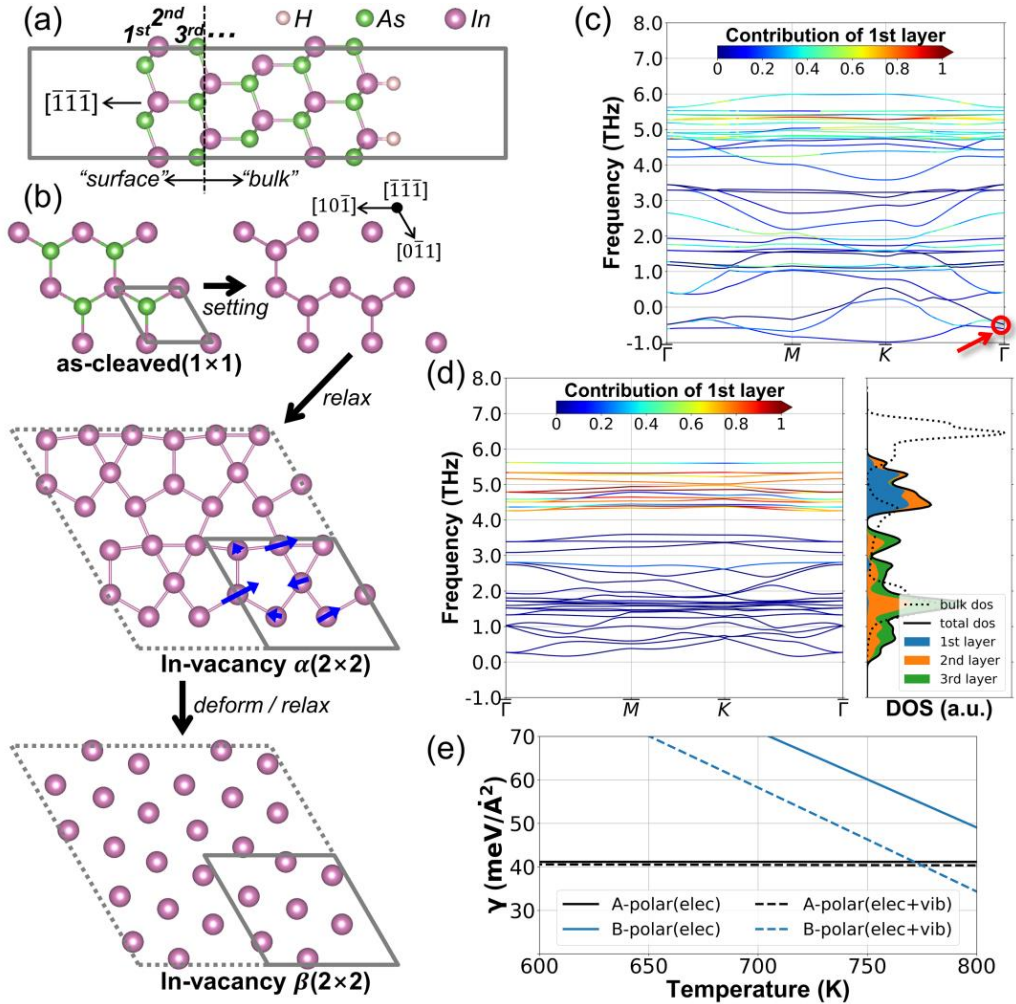

**Figure 3.** (a) Side view of the as-cleaved slab along the <111> direction of InAs. The bottom (111)A surface is passivated by hydrogens and the top three layers of the (111)B surface were selected for phonon evaluation. (b) Top view of the topmost bilayer (first and second layers) of (111)B as-cleaved(1×1), In-vacancy α(2×2), and In-vacancy β(2×2). The dotted parallelograms of the (2×2) reconstructions correspond to the supercell used for DFT force calculations, which comprises four unit cells (solid parallelogram). The set of blue arrows on a unit cell of the In-vacancy α(2×2) corresponds to the lowest imaginary mode at point $\bar{\Gamma}$. (c) The first-layer-projected band of the In-vacancy α(2×2). (d) The first-layer-projected band along with the layer-projected DOSs of the In-vacancy β(2×2) in comparison with bulk InAs (ZB, dotted line). (e) The electronic ($\gamma^{elec}$, solid line) and total surface energy ($\gamma^{elec} + \gamma^{vib}$, dotted line) of InAs(111)A In-vacancy(2×2) and (111)B In-vacancy β(2×2), which are labelled A-polar and B-polar, respectively, as a function of temperature at a fixed As-vapor pressure of $10^{-9}$ atm.

### 3.2. Adsorption: Oxygen on Cu(111)

The adsorption energy on a metal surface is widely used as an effective descriptor to predict catalytic activity. Based on a database of calculated adsorption energies, new catalysts with improved selectivity for complex surface reactions, such as the oxygen reduction reaction (ORR) and oxygen evolution reaction (OER), have been explored among a variety of metal species, alloy compositions, surface orientation, and so on [50,51]. In this section, the difference in vibration before and after the adsorption of oxygen on Cu(111) and its effects on energetic stability were examined to demonstrate how *InterPhon* can be applied to promote accurate adsorption-related calculations. This will contribute to various theoretical studies, including studies of deposition and growth from vapor as well as catalytic behavior.

The crystal structure of Cu is face-centered cubic (FCC, $Fm\bar{3}m$) which is centrosymmetric. Therefore, contrary to the ZB structure, the two opposite surfaces of (111) are equivalent; hence, a symmetric slab structure without hydrogen passivation was used. Figure 4a shows a side view of the slab along the <111> direction, while the corresponding top view is shown in Fig. 4b. For oxygen adsorption, two sites of Cu(111), FCC and hexagonal close-packed (HCP), were considered using an adsorption density of 0.25 ML, as shown in Fig. 4b. After optimizing each structure by relaxation, phonon calculation was carried out for the top three Cu layers and the adsorbed oxygen by setting the user-parameters of *InterPhon* as follows: **displacement** = 0.02; **enlargement** = "4 4 1" for pristine(1×1) and **enlargement** = "2 2 1" for O-FCC(2×2) and O-HCP(2×2); and **periodicity** = "1 1 0". Figure 4c-e shows the resulting phonon band structures along with the layer-projected DOSs of the pristine(1×1), O-FCC(2×2), and O-HCP(2×2), respectively, obtained by setting some of parameters of the 'plot_with_dos' method of 'Band' class as **band_option** = 'plain' and **dos_option** = 'line'. The DOSs projected on the Cu layers in each structure are summarized in Fig. 4f with that of bulk Cu (FCC, gray area) for comparison, where the label 'Cu layers' indicates the sum of the projection on Cu atoms in the top three layers (first, second, and third layers). It is evident that the phonon dispersion of the pristine surface is red-shifted but similar to that of bulk, which is attributed to the very similar bonding geometry but decreased number of bonds. In addition, the phonon modes in the high frequency range around 12 – 15 THz, which do not appear in the pristine surface, are found in the surface Cu layers after oxygen adsorption. The drastic change in vibrations is due to the formation of strong Cu–O bonds on the surface. It is also notable that the phonon frequency of the Cu layers in O-FCC(2×2) is higher than that in O-HCP(2×2), which is relevant to the 0.13 eV lower binding energy of O-FCC(2×2) than that of O-HCP(2×2).

These phonon DOSs were integrated to evaluate vibrational free energy ($F^{vib}$) [1,2,49], which is a constituent term in the change of Gibbs free energy by adsorption:

$$\Delta G_{adsorption} = G_{surf \cdot O} - G_{surf} - G_O = \Delta E^{elec}_{adsorption} + F^{vib}_{surf \cdot O} - F^{vib}_{surf} - \mu_{O(O_2)}, \quad (4)$$

where $G_{surf \cdot O}$ and $G_{surf}$ are the Gibbs free energy of the adsorbed surface and pristine surface, respectively; $\Delta E^{elec}_{adsorption}$ is the change in electronic energy by adsorption; $F^{vib}_{surf \cdot O}$ and $F^{vib}_{surf}$ are the vibrational free energy of the adsorbed surface and pristine surface, respectively; and $\mu_{O(O_2)}$ is the chemical potential of oxygen in vapor phase, which is available in thermochemical tables or can be calculated statistically under the ideal gas assumption [7]. Note that most previous DFT calculations on adsorption considered only the electronic adsorption energy, $\Delta E^{elec}_{adsorption}$. Recently, there have been attempts to evaluate the Gibbs free energy of adsorption within the following approximation [52]: the vibrations of surface region is not much affected by adsorption, so the free energy contribution from surface vibrations is negligible, $F^{vib}_{surf(\cdot O)} = F^{vib}_{surf}$. Under this approximation, only contributions from vapor species were considered, and the vibrational free energy of an adsorbate (*e.g.*, $F^{vib}_{(surf) \cdot O}$) was calculated using the discrete vibrational frequencies (*e.g.*, total three frequencies in the case of a monoatomic adsorbate) at point $\bar{\Gamma}$ [11,53]. Ignoring the vibrational contribution of the surface may be valid if the density of adsorbates is low. As the coverage and bonds between the adsorbate and surface increase, however, the contribution will no longer be negligible. Figure 4g shows the change in $F^{vib}$ by the oxygen adsorption contributed by the Cu(111) surface (corresponding to the red-highlighted terms in the full formulation). While the electronic energy of O-FCC(2×2) is lower than that of O-HCP(2×2) by 0.13 eV, the vibrational energy is higher in the FCC site. The difference in the vibrational energy between the FCC and HCP sites, originating from the higher frequency of O-FCC(2×2), increases as temperature increases and becomes comparable to the electronic energy difference. This is consistent with the following tendency observed in several bulk and surface systems: a structure with higher electronic energy (weaker bonding) shows lower vibrational energy, leading to a temperature- and entropy-driven phase transition [8,17,18,20,54].

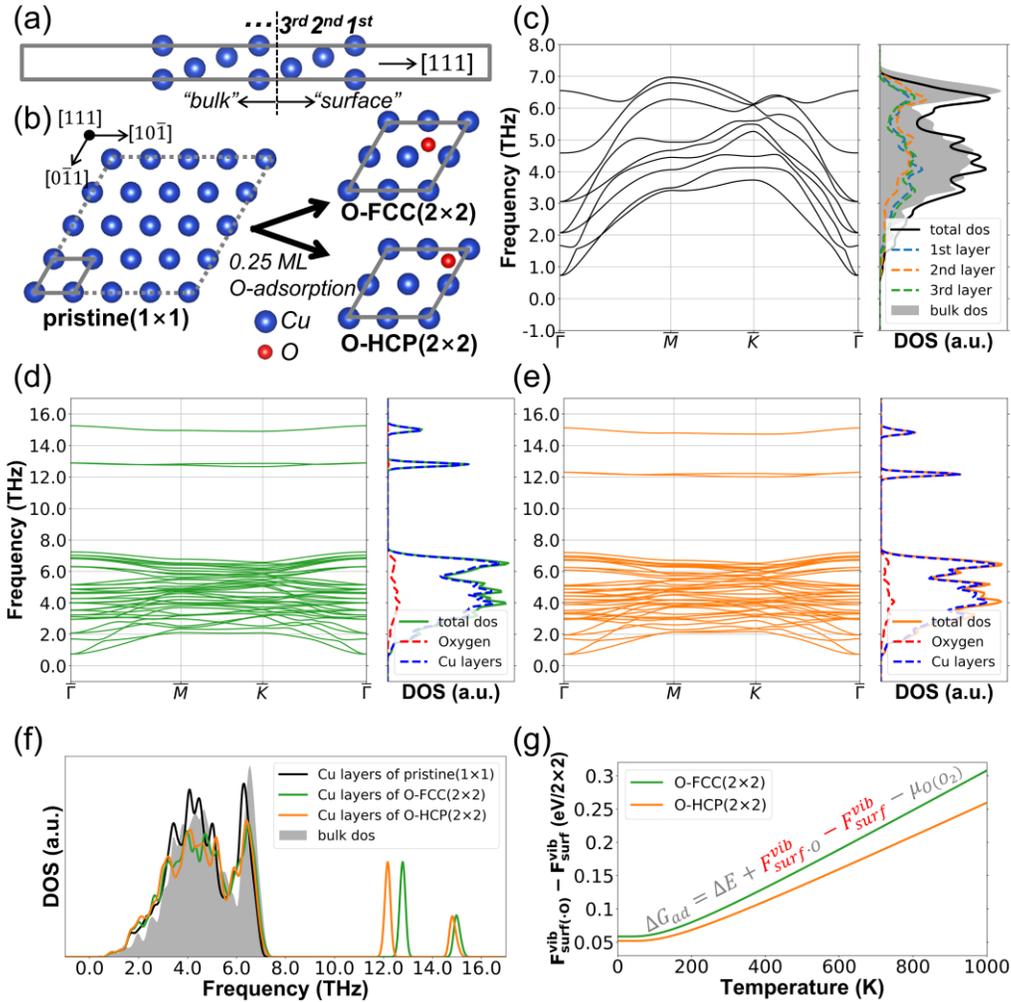

**Figure 4.** (a) Side view of a pristine slab along the <111> direction of Cu and (b) top view of the topmost layer (first layer) of the pristine (111) and that of (111) with adsorbed oxygen at the FCC and HCP sites with 0.25 ML coverage. The solid parallelograms indicate unit cells, while the dotted parallelogram of the pristine(1×1) corresponds to the supercell used for DFT force calculations, comprising sixteen unit cells (though not shown, supercells of the same size were used for the oxygen-adsorbed structures). The phonon band along with the layer-projected DOSs of the (c) pristine(1×1), (d) O-FCC(2×2), and (e) O-HCP(2×2). (f) Corresponding phonon DOSs projected to Cu atoms in the top three layers (first, second, and third layers) in comparison with bulk Cu (FCC, gray area). (g) Change in vibrational free energy of the Cu(111) surface (red-highlighted terms) by oxygen adsorption on the FCC and HCP sites.

### 3.3. Interface: Si(111)/GaAs(111)

The interfacial region between two different bulk phases is one of the most diverse and interesting material systems where a variety of interactions takes place. However, in numerous cases, structural models proposed by simple electronic energy calculations have failed to explain certain experimental observations, such as intermixing between GeTe/Sb$_2$Te$_3$ superlattices relevant to the resistance change of iPCM [55] and interface stoichiometry at epitaxial antiperovskite/perovskite heterostructures [56]. If an interfacial structure can successfully be described at the atomic scale by considering the contribution of vibrational energy, the interface control will be greatly facilitated, paving the way for unprecedented material design. In this section, using (111)-oriented Si/GaAs as an example, various types of interfacial structures and temperature-driven changes of those structures are investigated to demonstrate how *InterPhon* can be applied to predict interfacial structural transitions.

The crystal structure of Si is diamond (Fd$\bar{3}$m), which is centrosymmetric, while that of GaAs is ZB, which is non-centrosymmetric. As explained above with InAs, GaAs shows an intrinsic polarity along the <111> directions, exposing two inequivalent surfaces: Ga-terminated or A-polar (111)A surface; and the opposite As-terminated or B-polar (111)B surface. In contrast, the Si(111) surface is non-polar; thus, vertical growth of GaAs on a Si(111) substrate can be either A-polar or B-polar. Figure 5a shows a side view of the (111)-oriented Si/GaAs interfaces where three bonds between Si and Ga (or As) form. Throughout this section, this kind of interface will be referred to as type-I. On the other hand, at a type-II interface, only one bond between Si and Ga (or As) forms, as shown in Fig. 5b. The Si surface on the other side of the interface is passivated by hydrogen with one valence electron, while the GaAs surface on the other side of the interface is passivated by pseudo-hydrogen with 1.25 valence electrons for an A-polar surface and 0.75 valence electrons for a B-polar surface. Considering that Si is normally used as substrate, the in-plane lattices ($a_1$ and $a_2$) are fixed to those of Si, resulting in a biaxial compressive strain on GaAs (-4%). Relaxation was performed for the Si atoms in the interface layers and the whole GaAs layers, while the Si bulk layers were fixed. The corresponding top view of the two interface layers (1- and 2-layers of Si as well as I- and II-layers of GaAs) is shown in Fig. 5c,d. After optimization of each structure, phonon calculation was carried out for the three interface layers by setting the user-parameters of *InterPhon* as follows: **displacement** = 0.02; **enlargement** = "3 3 1"; and **periodicity** = "1 1 0". Figure 5e,f shows the resulting phonon band structures and DOSs of the type-I and type-II interfaces, respectively, with those of bulk Si (purple area) and bulk GaAs (gray area). For clarity, the phonon characteristics of the interfaces with Si–Ga bonds are represented by orange lines while those with Si–As bonds are represented by green lines.

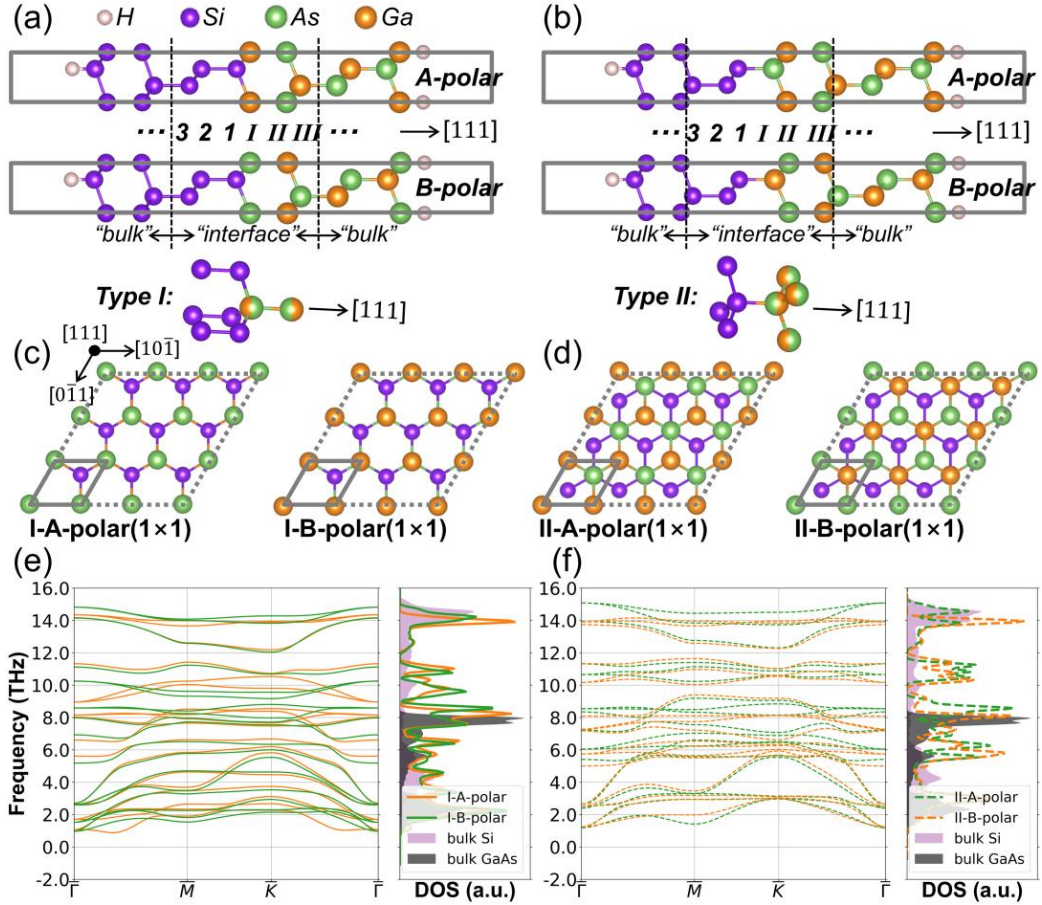

**Figure 5.** Side view of a slab along the <111> direction of Si/GaAs (a) type-I and (b) type-II interfaces. Both surfaces are passivated by hydrogens and the three layers along both Si and GaAs from the interface were selected for phonon evaluation. Top view of the two interface layers (1- and 2-layers of Si as well as I- and II-layers of GaAs) of (c) type-I and (d) type-II interfaces. The solid parallelograms indicate unit cells and the dotted parallelograms correspond to the supercell used for DFT force calculations, which accommodates nine unit cells. The phonon band structures along with the total DOSs of (e) type-I and (f) type-II interfaces, in comparison with bulk Si (purple area) and bulk GaAs (gray area). The orange (green) lines imply that Si–Ga (Si–As) bonds are formed at the interface.

The phonon DOSs of the interfaces were integrated to evaluate the phonon contributions to the free energy of interfaces [1,2,49]. Then, the vibrational free energy difference between interface and bulk, called the vibrational interface energy ($\gamma^{vib}$), was calculated:

$$\gamma^{vib} = \frac{F^{vib}_{interface} - N_{Si} \cdot F^{vib}_{Si(bulk)} - N_{Ga} \cdot F^{vib}_{GaAs(bulk)} - (N_{As} - N_{Ga}) \cdot F^{vib}_{As(GaAs)}}{A}, \quad (5)$$

where $F^{vib}_{interface}$, $F^{vib}_{Si(bulk)}$, $F^{vib}_{GaAs(bulk)}$, and $F^{vib}_{As(GaAs)}$ are the vibrational free energies of the interface, bulk Si, bulk GaAs, and As in bulk GaAs, respectively; $N_{Si}$, $N_{Ga}$, and $N_{As}$ are the numbers of Si, Ga, and As atoms in the interfacial region, respectively; and $A$ is the interfacial area. Figure 6a shows the $\gamma^{vib}$ of each type of interface. Interestingly, the contribution of vibrational energy to interface energy is negative for both type-I interfaces, while it is positive for the type-II interfaces. This stark contrast is attributed to the higher phonon frequency in type-II interfaces than type-I interfaces, as shown in Fig. 6b, where the phonon DOSs are compared between different types with the same interface bond, either Si–Ga (type-I-A-polar and type-II-B-polar) or Si–As (type-I-B-polar and type-II-A-polar). As in the previous examples, as temperature increases, the role of vibrational contribution becomes more important as the difference in $\gamma^{vib}$ between the interfaces increases.

Figure 6c shows the relative type-II vs. type-I interface energy ($\Delta\gamma = \gamma_{II} - \gamma_{I}$) at a fixed As-vapor pressure of $10^{-5}$ atm for both cases: interface energy evaluated by (i) only electronic contribution ($\gamma = \gamma^{elec}$) and (ii) both electronic and vibrational contributions ($\gamma = \gamma^{elec} + \gamma^{vib}$). The calculation method of how to obtain the temperature-pressure dependent interface energy are the same as the method for surface energy introduced with Fig. 3e [21,22]. Note that the relative electronic interface energy within the same polarity ($= \gamma^{elec}_{II-A/B-polar} - \gamma^{elec}_{I-A/B-polar}$) is the best that could possibly be obtained, since the absolute energy of a strained polar surface or interface cannot be calculated using the current levels of DFT simulation [57,58]. By definition, the positive relative interface energy means that the type-I interface is more stable than the type-II interface. The results evaluated only by electronic contribution are represented by solid lines, while those reflecting the sum of electronic and vibrational contribution are shown with dotted lines. In particular, Fig. 6c demonstrates that the type-I interface is more stable than type-II only for the B-polar interface. In addition, the stable temperature ranges of the I-B-polar increase from ~670 to ~740 K by considering the vibrational contribution. Finally, the temperature-pressure boundary between the stability region of the I-B-polar and II-B-polar interfaces are shown in Fig. 6d, which was obtained by extending the investigation of Fig. 6c to arbitrary pressure conditions. For the purposes of comparison, the region in which I-B-polar is predicted to be stable only by $\Delta\gamma^{elec}$ is also represented in dark blue. This prediction is remarkably well-matched with the results of previous experiments on the integration of GaAs nanowires (NWs) on Si substrate: pretreatment of the initial Si(111) surface with an As-source of selective-area metal-organic vapor phase epitaxy (SA-MOVPE, ~$10^{-5}$ atm) at low temperature (673 K) modifies non-polar (111) to B-polar (111)B. Followed by a gradual temperature increase to growth temperature (1023 K) and Ga-source injection, the surface pretreatment eventually facilitates the growth of vertically aligned GaAs NWs on a Si(111) substrate, achieving 100% yield [59], since GaAs NWs preferentially grow along the <111>B rather than <111>A direction [47]. According to our prediction, Si–As bonding at a type-I interface leads to vertical B-polar directionality (bottom panel of Fig. 5a), while Si–As bonding at a type-II interface leads to vertical A-polar directionality (top panel of Fig. 5b). Therefore, taking into account the vibrational contribution, the systematic calculations on the Si/GaAs interface clearly explain why pretreatment of the Si(111) substrate by an As-source had to be carried out under low-temperature conditions to promote vertically aligned growth of B-polar GaAs NWs.

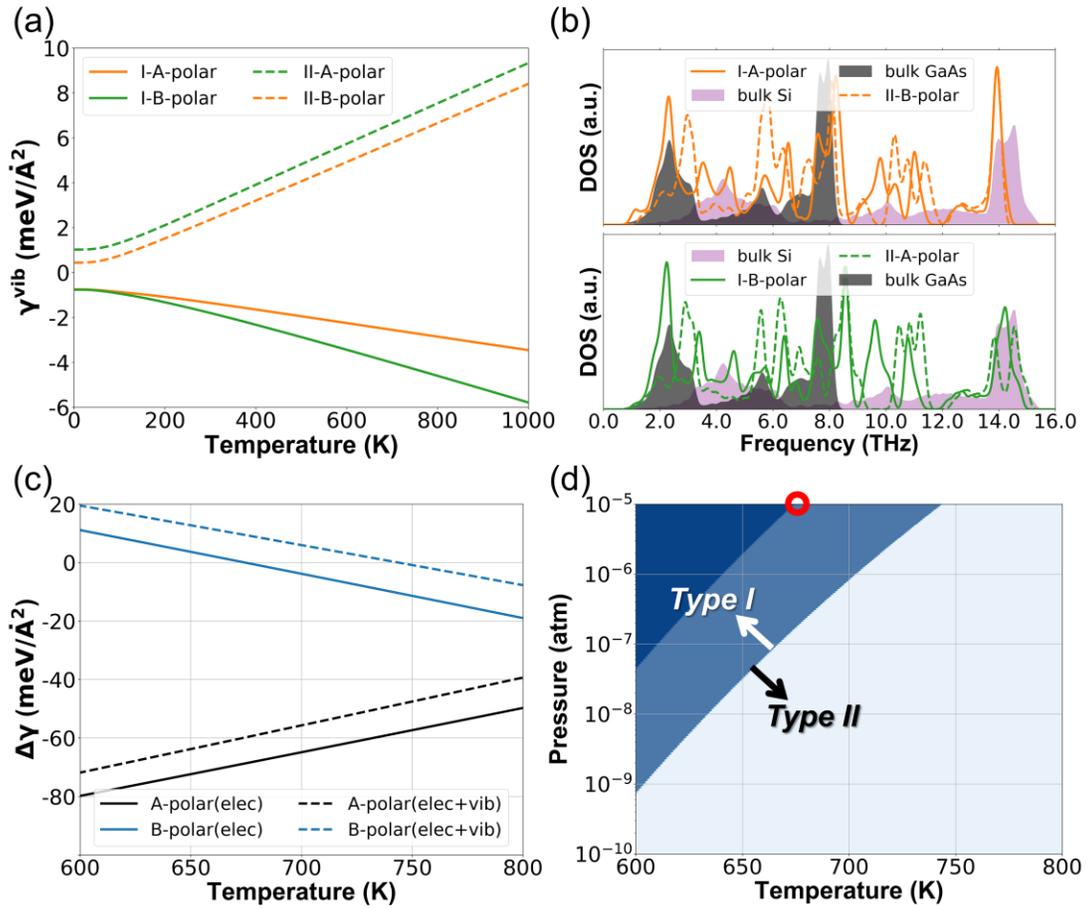

**Figure 6.** (a) The vibrational interface energy ($\gamma^{vib}$) of type-I (solid lines) and type-II (dotted lines) interfaces. (b) The total DOSs of the interfaces with Si–Ga bonds (top panel) and those with Si–As bonds (bottom panel). (c) The relative interface energy of type-II compared to type-I interfaces

($\Delta\gamma = \gamma_{II} - \gamma_I$) obtained using electronic interface energy ($\gamma = \gamma^{elec}$, solid lines) and total interface energy ($\gamma = \gamma^{elec} + \gamma^{vib}$, dotted lines) at a fixed As-vapor pressure of $10^{-5}$ atm. (d) The temperature-pressure boundary where the stable interface type changes from I-B-polar ($\gamma_{I-B-polar} < \gamma_{II-B-polar}$) to II-B-polar ($\gamma_{I-B-polar} > \gamma_{II-B-polar}$) predicted using the total interface energy. The dark blue area at the top left corner corresponds to the region where I-B-polar is predicted to be stable by the electronic interface energy. The red circle marks the experimental condition (673 K, $10^{-5}$ atm) [59] under which pretreatment of Si(111) substrate by an As-source was carried out.

## 4. Conclusions

To facilitate theoretical studies on interfacial phonons, a Python package (*InterPhon*) connected to the 3D framework of available DFT codes was developed, which supports a set of methods to efficiently extract localized lattice vibrations with high-level user interfaces. The *InterPhon* strategy of user-defined interfacial region and user-defined periodicity was validated showing that the phonons of atom layers far from the interface gradually converge to the bulk phonons. Furthermore, the instructions of design architecture and package workflow were provided so as to be ready for a potential adaptation to unexpected problems and other computational projects. The usability of this package was thoroughly demonstrated through its application to representative surface, adsorption, and interface examples with complete availability of data provenance. In the first example, the separate sets of phonon calculations on the two inequivalent polar surfaces, InAs (111)A and (111)B, allows us to investigate the difference in phonon properties between the opposite surfaces, which might be impossible without limiting the range of phonon calculations. Meanwhile, the reconstruction of InAs (111)B In-vacancy β(2×2), which is far more stable than In-vacancy α(2×2) obtained by DFT relaxation, was found by the guidance of imaginary phonon mode of the unstable In-vacancy α(2×2) surface reconstruction. In the second example of oxygen adsorption on Cu(111), the surface phonon differences between different adsorption configurations are investigated showing that a configuration with lower electronic energy has higher vibrational free energy. This finding implies the possibility of temperature-induced changes in a stable configuration, usually called entropy-driven transition. In the third example of Si/GaAs interfaces, distinct phonon patterns of different interface structures enable the predictions of interface transitions depending on the temperature-pressure condition, of which validity was confirmed by comparison with a relevant experiment on the growth of GaAs NWs on Si substrate.

Following the examples of *InterPhon* application, the package will be of great help in a variety of theoretical studies on material interactions, such as material growth on surfaces, catalytic activity of adsorption, and correlated phenomena at interfaces, by providing opportunities to solve numerous cases of mismatch between the structures predicted by simple electronic energy calculations at 0 K and the experimental observations at finite temperature. The implementation of this efficient method and its full automation in conjunction with the currently available framework will facilitate automatic and massive computations of interfacial phonons, previously deemed inaccessible, and help to elucidate a variety of unexplored interfacial dynamics.

# Author contributions

In Won Yeu: Conceptualization, Methodology, Formal analysis, Investigation, Visualization, Software, and Writing-Original Draft. Gyuseung Han: Validation and Writing - Review & Editing. Kun Hee Ye: Validation and Writing - Review & Editing. Cheol Seong Hwang: Conceptualization and Writing - Review & Editing. Jung-Hae Choi: Conceptualization, Resources, Validation, Supervision, Funding acquisition, Project administration and Writing - Review & Editing.

# Conflicts of interest

There are no conflicts to declare.

# Acknowledgements

This work was supported by a National Research Foundation of Korea (NRF) grant funded by MSIT [2020R1A2C2003931], and by the Institutional Research Program of the Korea Institute of Science and Technology (KIST) [2E30410]. The authors also acknowledge the computational resources provided by the "HPC Support" Project, supported by MSIT and NIPA, Korea.

# Appendix A. Supplementary material

Details of phonon formalism using FDM, validation of symmetry functionality, *InterPhon* package architecture, method to define the interfacial region and convergence test, and calculation details for all of the results in this work.

## Appendix B. Vector and matrix notations

The coordinates of an atom are defined by three Cartesian components:

$$\boldsymbol{R}_{ls} = \begin{pmatrix} R_{lsx} \\ R_{lsy} \\ R_{lsz} \end{pmatrix}, \qquad (B1)$$

where $l$ and $s$ labels indicate the lattice and basis, respectively.

Displacement vector of the atom located at $\boldsymbol{R}_{ls}$ along an arbitrary direction $\boldsymbol{u}_1$ is represented by a column vector:

$$\boldsymbol{u}_{ls1} = \begin{pmatrix} u_{lsx} \\ u_{lsy} \\ u_{lsz} \end{pmatrix}. \qquad (B2)$$

Force vector of the atom at $\boldsymbol{R}_{ls}$ induced by the displacement of the atom at $\boldsymbol{R}_{l's'}$ along $\boldsymbol{u}_1$ direction is written as:

$$\boldsymbol{F}_{ls,l's'1} = \begin{pmatrix} F_{lsx,l's'1} \\ F_{lsy,l's'1} \\ F_{lsz,l's'1} \end{pmatrix}. \qquad (B3)$$

Force constant matrix between the two atoms at $\boldsymbol{R}_{ls}$ and $\boldsymbol{R}_{l's'}$ is represented by 3×3 matrix:

$$\boldsymbol{K}_{ls,l's'} = \begin{pmatrix} K_{lsx,l's'x} & K_{lsx,l's'y} & K_{lsx,l's'z} \\ K_{lsy,l's'x} & K_{lsy,l's'y} & K_{lsy,l's'z} \\ K_{lsz,l's'x} & K_{lsz,l's'y} & K_{lsz,l's'z} \end{pmatrix} = - \begin{pmatrix} \frac{\partial F_{lsx}}{\partial u_{l's'x}} & \frac{\partial F_{lsx}}{\partial u_{l's'y}} & \frac{\partial F_{lsx}}{\partial u_{l's'z}} \\ \frac{\partial F_{lsy}}{\partial u_{l's'x}} & \frac{\partial F_{lsy}}{\partial u_{l's'y}} & \frac{\partial F_{lsy}}{\partial u_{l's'z}} \\ \frac{\partial F_{lsz}}{\partial u_{l's'x}} & \frac{\partial F_{lsz}}{\partial u_{l's'y}} & \frac{\partial F_{lsz}}{\partial u_{l's'z}} \end{pmatrix}, \qquad (B4)$$

which describes the proportionality between displacements and forces:

$$F_{ls,l's'\mathbf{1}} = -K_{ls,l's'}u_{l's'\mathbf{1}}. \tag{B5}$$

The whole force constant matrix becomes 3*N*×3*M* matrix where *N* and *M* are the number of interfacial atoms in supercell and unit cell, respectively.

A component of the total force vector acting on atom $\mathbf{R}_{ls}$ is evaluated by sums over forces contributed by all of the atomic displacements:

$$F_{ls\alpha} = -\sum_{l's'\beta} K_{ls\alpha,l's'\beta}\, u_{l's'\beta}, \tag{B6}$$

where the labels of $\alpha$ and $\beta$ indicate one of the three Cartesian components. This is the same expression as Equation S3 derived from the Newtonian equation of motion.

## Appendix C. Symmetry operation

An affine mapping ($\mathbf{W}$, $\mathbf{w}$) transforms $\mathbf{R}_{ls}$ into an image point:

$$\mathbf{R}_{\widetilde{ls}} = \mathbf{W}\mathbf{R}_{ls} + \mathbf{w}, \tag{C1}$$

where $\mathbf{W}$ is a rotation part represented by 3×3 matrix and $\mathbf{w}$ is a translation part represented by 3×1 column vector:

$$\mathbf{W} = \begin{pmatrix} W_{1,1} & W_{1,2} & W_{1,3} \\ W_{2,1} & W_{2,2} & W_{2,3} \\ W_{3,1} & W_{3,2} & W_{3,3} \end{pmatrix}; \quad \mathbf{w} = \begin{pmatrix} w_1 \\ w_2 \\ w_3 \end{pmatrix}. \tag{C2}$$

The image point $\mathbf{R}_{\widetilde{ls}}$ corresponds to the coordinate of another atom if ($\mathbf{W}$, $\mathbf{w}$) belongs to symmetry operations of the crystal. In particular, applying $\mathbf{W}$ on $\mathbf{R}_{ls}$ not only changes the coordinates but also rotates relevant displacement and force vector fields. If a vector $\boldsymbol{u_1}$ is rotated onto a vector $\boldsymbol{u_2}$ by $\boldsymbol{W}$, applying $\boldsymbol{W}$ on $\boldsymbol{u_{ls1}}$ and $\boldsymbol{F_{ls,l's'1}}$ results in:

$$\boldsymbol{u_{\widetilde{ls}2}} = \boldsymbol{W}\boldsymbol{u_{ls1}}; \quad \boldsymbol{F_{\widetilde{ls},\widetilde{l's'}2}} = \boldsymbol{W}\boldsymbol{F_{ls,l's'1}}. \tag{C3}$$

## Appendix D. Derivation of Equation 3

From Equation C3:

$$\boldsymbol{F_{\widetilde{ls},\widetilde{l's'}2}} = \boldsymbol{W}\boldsymbol{F_{ls,l's'1}}$$

Insert the proportionality relation between displacement and force vectors (Equation B5):

$$\boldsymbol{K_{\widetilde{ls},\widetilde{l's'}}}\boldsymbol{u_{\widetilde{ls}2}} = \boldsymbol{W}\boldsymbol{K_{ls,l's'}}\boldsymbol{u_{ls1}}$$

Multiply on the left side with $\boldsymbol{W^{-1}}$ and replace $\boldsymbol{u_{\widetilde{ls}2}}$ by $\boldsymbol{W}\boldsymbol{u_{ls1}}$:

$$\boldsymbol{W^{-1}}\boldsymbol{K_{\widetilde{ls},\widetilde{l's'}}}\boldsymbol{W}\boldsymbol{u_{ls1}} = \boldsymbol{K_{ls,l's'}}\boldsymbol{u_{ls1}}$$